\documentclass[twocolumn,showpacs,preprintnumbers,amsmath,amssymb, aps,
prb,superscriptaddress]{revtex4}
\usepackage{graphicx}
\usepackage{times}
\usepackage{dcolumn}
\usepackage{bm}
\usepackage{array}
\usepackage{hyperref}
\usepackage{amsmath}
\usepackage{color}

\begin{document}
\title{Dynamically reconfigurable  directionality of plasmon-based single photon sources}
\author{Yuntian Chen}
\affiliation{ DTU Fotonik, Department of Photonics Engineering, \O
rsteds Plads, Building 343, DK- 2800 Kgs. Lyngby, Denmark.}
\affiliation{Center for Nanophotonics, FOM Institute for Atomic and
Molecular Physics (AMOLF),
 Science Park 104, 1098 XG Amsterdam, The Netherlands}
\author{Peter Lodahl}
\affiliation{ DTU Fotonik, Department of Photonics Engineering, \O
rsteds Plads, Building 343, DK- 2800 Kgs. Lyngby, Denmark.}
\author{A. Femius Koenderink}
 \email{fkoenderink@amolf.nl}
\affiliation{Center for Nanophotonics, FOM Institute for Atomic and
Molecular Physics (AMOLF),
 Science Park 104, 1098 XG Amsterdam, The Netherlands}
\date{\today}
\begin{abstract}
We propose a plasmon-based reconfigurable antenna to controllably
distribute emission from single quantum emitters in spatially separated
channels. Our calculations show that crossed particle arrays can
split the stream of photons from a single emitter into multiple
narrow beams. We predict that beams can be switched on and off by
switching host refractive index. The design method is based on engineering the dispersion
relations of plasmon chains and is generally applicable to
traveling wave antennas. Controllable photon delivery has
potential applications in classical and quantum communication.
\end{abstract}
 \maketitle

Controllably and efficiently extracting photons from single quantum
emitters into a well-defined set of modes is a holy grail for
quantum optics, optical quantum computation, as well as single
molecule spectroscopy. The conventional approach is to place the
emitter inside a high finesse ultrasmall cavity, such as a
micropillar \cite{GerardPRL1998,PeltonPRL2002}, microsphere or
toroid \cite{ArmaniNature2003}, or photonic crystal cavity
\cite{YoshieNature2007}. Alternatively, several groups have started
to pursue plasmonic systems for quantum optics. \cite{chang053002}
By virtue of the large interaction strength of free electrons in
noble metals with photons at optical frequencies, plasmon polaritons
offer very tight field confinement over large frequency bandwidths.
In addition to applications in subwavelength optoelectronics and
near-field sensors, \cite{Smolyaninov2005} plasmonics hence offers
rich perspectives for quantum optics with single plasmons
\cite{chang053002}, and for novel broadband single-photon sources
based on plasmon antennas. For instance, several researchers
recently proposed that broadband highly directional single-photon
sources can be made using plasmon particle array antennas that mimic
directional radio frequency antennas.
\cite{Hofmann2007,Koenderink2009Nanoletters} We also note that
ultrafast plasmonic  phenomena and all-optical plasmon modulators
were studied recently. \cite{Macdonald2009}

In a quantum network in which several localized qubits interact
via emission of photons, one would desire reconfigurable
coupling between nodes in the network of qubits. By analogy to
radio-wave antennas, one might expect that plasmon antennas
used to control emitters can be reprogrammed with ease to
arbitrarily steer beams. However, programmable radio-wave
antennas use methods inaccessible to plasmonics, as they usually
use individual phase control over many active elements. In this
paper we propose a new strategy to obtain control of
reconfigurable plasmon antennas for single emitters.  Our method
rests on controlling the dispersion relation of guided modes in
each part of a multi-arm  traveling wave plasmon antenna by
switching the refractive index of the surrounding medium. Intuitively,
the large bandwidth of plasmonic antennas implies that larger
index changes are needed to switch than in high Q dielectric
cavities. We show that an effective reconfigurable switch can be
reached with host index changes that are achievable with liquid crystals. \cite{ElKallassi2007JOSAB}

\begin{figure}
\includegraphics[width=0.8\columnwidth]{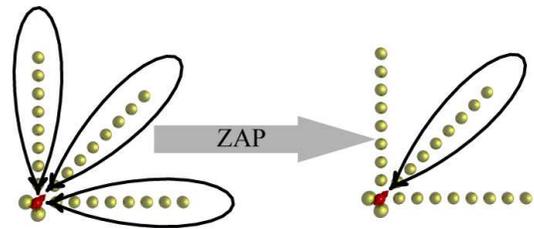}
\caption{\label{cartoon_switch} Sketch of our reconfigurable
nano-antenna concept to control single emitters. We consider
a single emitter (red dipole) embedded in a set of linear plasmon
antennas (metal particles in yellow)  that intersect
at the emitter. In its unswitched state (left), such an
antenna funnels spontaneous emission into different beams.
The beams can be switched on and off (indicated by ``ZAP") at will
by modifying the particle or host material dynamically.}
\end{figure}

\begin{figure}\centering
\includegraphics[scale=0.5]{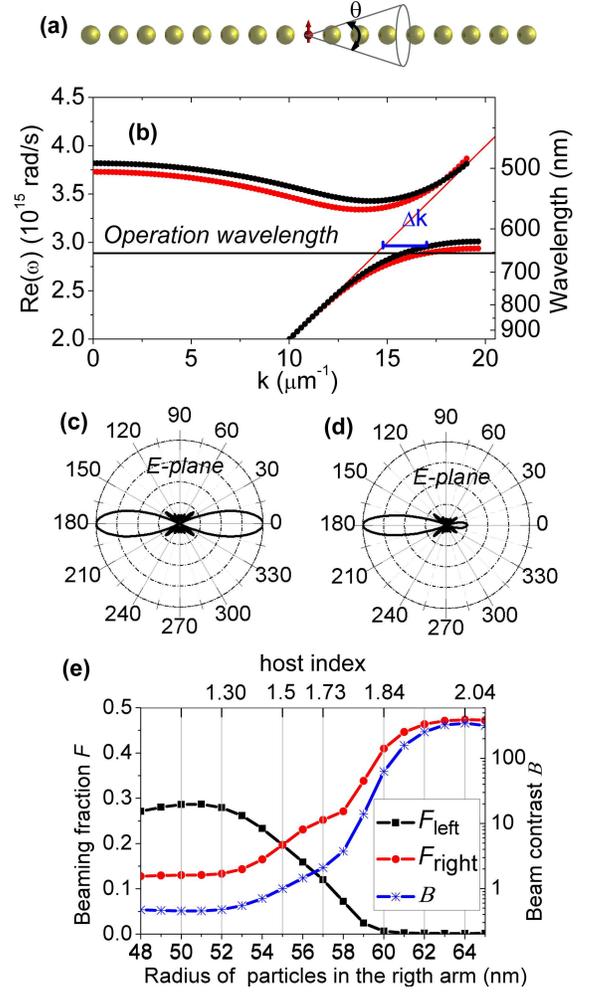}
\caption{\label{symmetric} (a) Sketch of  the emitter-antenna
geometry consisting of two identical arms. We consider a single
emitter placed in the middle of a 160 nm gap between the antenna
arms, oriented perpendicular to the antenna axis.  (b) The black
(red) curve represents the dispersion relation for the transverse
plasmon modes of an infinite Ag particle chain in glass (pitch d=160
nm, particle size R=55  (58) nm). The black horizontal line
indicates the operation wavelength at 652 nm. (c/d) Emission pattern
at $\lambda= 652$ nm for a single emitter embedded in a linear array
antenna, with 8 silver particles to each side. In (c) both arms are
equal (R=55 nm). In (d)  the right arm has R=58 nm. (e) Radius/host
index (top axis) dependence of the beaming fraction and beam
contrast, assuming collection in a cone of width $\sin\theta=0.32$.}
\end{figure}

We consider multi-beam antennas that split the stream of photons
emitted by a single emitter into several channels, as shown in
Fig.~\ref{cartoon_switch}, each corresponding to a narrow beam of
$<$ 30 degree full width at half maximum.
\cite{Hofmann2007,Koenderink2009Nanoletters} We explore the
possibility of dynamically switching on and off each beam at will,
for instance by controlling the refractive index surrounding the
antenna. We envisage that such a dynamically reconfigurable
multi-beam antenna can be useful in quantum optics, to controllably
couple a local qubit to a select number of other qubits. First, let
us consider how the multi-beam antenna works in its unswitched
state. Following a proposal by Li \emph{et al.}\cite{Li2009}, we
propose that a multi-beam antenna with $N$ beams can be made by
combining  $N$ antenna arms that each consist of a linear array of
metal particles, and essentially act like Yagi-Uda type antennas at
optical frequencies. Recent reports have shown that such antennas
can force single emitters to emit into a narrow beam over a broad
bandwidth that is demarcated on the blue edge by an abrupt cut-off.
The cut-off wavelength depends on antenna geometry
\cite{Koenderink2009Nanoletters}. The physics can be understood by
considering a Yagi-Uda antenna as a traveling wave antenna, the
behavior of which is governed by the dispersion relation for an one
dimensional infinite plasmon chain. \cite{Koenderink2006prb} When
the emission frequency is tuned to the lower dispersion branch, the
emitter decays into a plasmonic mode bound to the antenna, and with
a wave vector beyond the light line, see Fig.~\ref{symmetric}-(b).
The finite antenna length causes efficient out-coupling of this
mode, which hardly radiates in the case of infinite plasmon chains.
For a linear plasmon particle array of length $L$, momentum
conservation is only defined within $\triangle k\approx \pi/L$. This
determines the cut-off wavelength of efficient beaming. The
wavelength at which the dispersion relation deviates more than
$\triangle k$ from the light line, marked by the blue bar in
Fig.~\ref{symmetric}-(b), corresponds to the cut-off wavelength. If
the operation wavelength denoted by $\lambda_{op}$, is longer than
the cut-off wavelength, the  plasmon chain acts as a directional
antenna for single-photon emission. If $\lambda_{op}$ is shorter
than the cut-off wavelength, the emitter decays into dark plasmons.
\cite{Koenderink2009Nanoletters} Importantly, the  cut-off is very
sharp and occurs within a few nanometer spectral bandwidth.
\cite{Koenderink2009Nanoletters} Such abrupt on/off behavior is
essential for optical switching of plasmon antennas.

As a first example, we study the coupling of a single emitter to an
antenna with two identical arms,  consisting of silver spheres
\footnote{We use tabulated constants from E. D. Palik, Handbook of
optical constants of solids II (Academic Press, 1991).} (radius
$R=55$ nm) , arranged in a linear array with pitch of $d=160$ nm,
shown in Fig.~\ref{symmetric}-(a). The array is embedded in glass
($n=1.5$) and the dipole emitter is transverse to the arrays. The
real part of  the corresponding infinite chain dispersion relation
for the transverse mode, (black curve in Fig.~\ref{symmetric}-(b)),
is calculated from a point-dipole model. \cite{Koenderink2006prb}
Since both arms are identical, they have exactly the same dispersion
relation, and the emitted photon is split into two identical beams.
As in the case of a single Yagi-Uda antenna, the beams have a full
width at half maximum of 30 degrees, as calculated using ``MESME".
``MESME" is an exact electrodynamic multiple scattering multipole
expansion method developed by F. J. Garc\'{\i}a de Abajo for
rigorously solving Maxwell's equations for finite clusters of
scatters. \cite{Abajoprbprl1998,Koenderink2009Nanoletters} The fact
that we choose a linear antenna ($180^0$  between arms) is not
essential: we obtain similar splitting into two beams for
perpendicular arms.

We consider how much perturbation is required to switch one of the
two  beams off. Two facts are immediately obvious: First, since we
start with a symmetric antenna, we require an asymmetric
perturbation to switch only one of the beams. Second, we expect a
dramatic change in emission pattern only if the perturbation shifts
the cut-off wavelengths through $\lambda_{op}$. Therefore
$\lambda_{op}$ is chosen close to the cut-off wavelength. Before
focusing on a specific switching mechanism, we note that the key
parameter that determines the dispersion is the polarizability
$\alpha$ of each particle. In the electrostatic approximation we
have $\alpha =3V (\varepsilon-n^2)/(\varepsilon +2n^2)$, with
particle volume $V=4 \pi R^3/3$, host index $n$ and  metal
dielectric constant $\varepsilon$. To obtain a first estimate for
the amount $\triangle \alpha$ needed to shift the dispersion
sufficiently, we vary $\triangle \alpha$ through $\triangle R$, even
though this may not be physically realizable in a dynamical manner.
We discuss realistic implementations below. We find that at fixed
pitch and host index, the dispersion red shifts as particle size
increases, c.f the red curve in Fig.~\ref{symmetric} (b). When the
particle size is increased from $R=55$ nm to $R=58$ nm, the shift
amounts to $\sim 20$~nm, which moves the cut-off wavelength through
$\lambda_{op}$. We therefore expect a dramatic change in radiation
pattern. Indeed the calculation (Fig.~\ref{symmetric}-(c-d)) shows
that a single  beam remains  from the unswitched arm, and
disappearance of the beam from the switched arm.

\begin{figure}\centering
\includegraphics[scale=0.5]{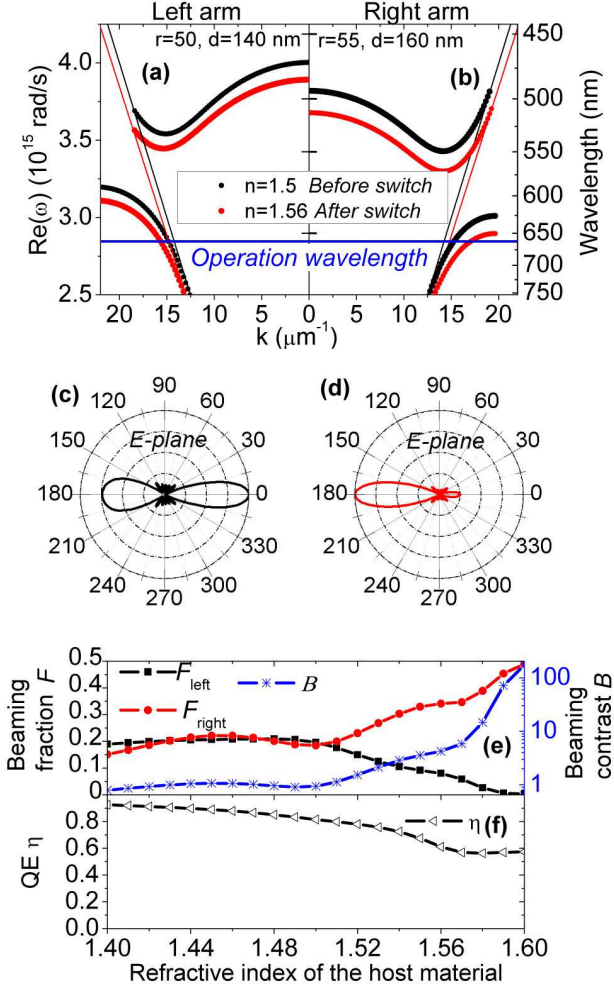}
\caption{\label{asymmetric} Results for an asymmetric two beam
antenna, with radius $R=50$~nm, pitch $d=140$~nm  (left arm) and
$R=55$~nm, $d=160$~nm in the right arm. (a/b) shows the dispersion
relation for transverse plasmons for each arm before (black curve)
and after the switch (red curve). (c/d) Emission pattern for a
single emitter in the antenna (both arms with 8 particles) before
(c) and after (d) switching host index from 1.5 to 1.56. (e)  Host
index dependence of the beaming fraction and beam contrast at
$\lambda= 662$ nm. (f) Quantum efficiency versus the variation of
host index at $\lambda= 662$ nm.}
\end{figure}

In order to quantify the quality of the switching behavior,  we
define two figures of merit. The first figure of merit called the
beaming fraction $F$, quantifies how much of the total emitted power
is emitted into the left arm and right arm, respectively
${F_{{\text{left/right}}}} = {\int_{_{{\text{(}}{\Omega
_0}{\text{,left/right)}}}} {Pd\Omega}}/{\int_{_{(4\pi )}} {Pd\Omega
} }$, where $P$ is the power radiated per solid angle. We define a
solid angle $\Omega_0$, which we take to correspond to a numerical
aperture $\texttt{NA}=sin(\theta)$, that one would use to collect
the radiation of each beam in practice. The second figure of merit
called the beam contrast $B  =
{{F_{{\text{left}}}}}/{{F_{{\text{right}}}}}$ quantifies the on/off
contrast, and is defined as the brightness contrast between the two
arms. We plot both figures of merit in Fig.~\ref{symmetric}-(e) for
different magnitudes of the perturbation of the right hand arm of
the antenna. At $R=55$ nm, both arms are equal and carry equal
amounts of energy ($B=1$). For a fixed $\texttt{NA}=0.32$ one would
collect a fraction of $\sim 20\%$ of emitted power in each beam. For
particle size R=58 nm in the right hand beam, the right  beam is
strongly reduced to below $\sim 3\%$. At the same time the left beam
gains a factor two in brightness. The contrast between the beams
hence shifts from $B=1$ to several hundred. In order to translate
the required $\triangle R$ back to  a physically realizable switch,
we note that $\Delta R/R \sim 10\%$. We hence conclude that a
two-beam antenna with identical arms can be reconfigured provided
one finds a way to change the polarizability of particles in one arm
of the antenna by $30\%$. Since the only feasible method to change
polarizability is to change the host index, we convert $\triangle
\alpha$ into a required change in host index (top axis in
Fig.~\ref{symmetric} (e)). An immense change from $n=1.5$ to
$n=1.85$ would be required, which is unachievable in any practical
material.  We conclude that prospects for switching are dim when one
starts out from antennas that are symmetric.

\begin{figure}\centering
\includegraphics[scale=0.42]{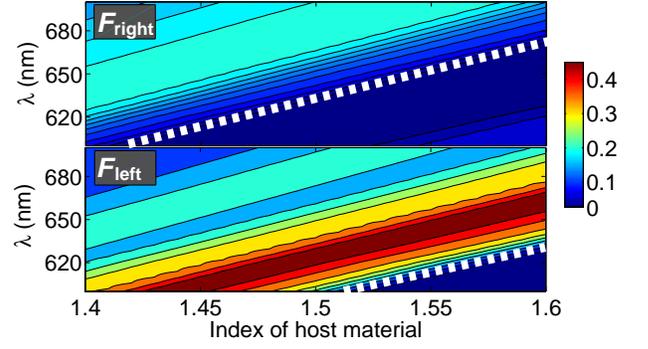}
\caption{\label{controuplot} Contour plot of beaming  fraction $F$
versus host index and wavelength. The dashed white lines denote the
cut-off wavelengths for the directional emissions. In a 30 nm band
to the red of the cutoff of the right arms, both beams have
comparable power. In between the cutoffs, the right beam is off, and
the left beam is brighter.}
\end{figure}

To overcome these challenges, we now consider a  two-beam antenna
that consists of different arms to begin with. Due to the asymmetric
geometry in which particle size and pitch are chosen smaller ($R=50$
nm, $d=140$ nm) in the left arm already at the fabrication stage,
the dispersion relation for the two arms are shifted already in the
unswitched case, shown in Fig.~\ref{asymmetric} (a). In the right
hand branch, $\lambda_{op}$ is much closer to the cut-off than in
the left branch. This yields the possibility of switching just one
beam off selectively by a homogenous switch in host material index,
provided that the cut-off of the right hand branch shifts beyond
$\lambda_{op}$, while $\lambda_{op}$ remains to the red side of the
cut-off in the left arm.  Figure~\ref{asymmetric} (c) shows that in
the unswitched ($n=1.5$) state, the emission from the emitter is
indeed split into two beams. Due to the intrinsic asymmetry in
geometry, both beams do not have  equal angular width, although they
carry comparable power. As the host index is raised from 1.5 to
1.56, the dispersion curves bend further away from the light line,
shifting the cut-off wavelength in both branches to the red
(Fig.~\ref{asymmetric} (b)).  Figure~\ref{asymmetric} (d) confirms
that the left beam remains, while the right beam switches off, in
accordance with the shift in cut-off wavelengths.

Figure~\ref{asymmetric} (e) allows us to assess the figures of merit
of the proposed switch, assuming $\lambda_{op}=662$ nm. We plot the
dependence of the beaming fraction, and the beam contrast as a
function of host index. While the beam contrast is approximately
equal at $n=1.5$, we see a marked contrast between left and right
beams at $n>1.56$. Hence we conclude that beams in a multi-beam
antenna can be switched at will with a manageable index change,
provided we carefully choose the dispersion relation for each
antenna branch. The role of the dispersion relation is further
confirmed in Fig.~\ref{controuplot}, which shows the beaming
fraction for the two arms as a function of both operation wavelength
and host index. There is a band throughout which each arm shows
pronounced directional emission with approximately equal power. To
the blue side of the cut-off of the right hand arm, the right hand
beam has very low intensity, and the intensity of the left arm is
increased by a factor of  ~2. To the blue of the cut-offs of both
waveguides, neither arm generates a bright and directional beam.
This result shows that the switch can be optimized for any
$\lambda_{op}$ in  a wide band, by antenna geometry and host index,
and that due to the sharp antenna cut-off only modest index changes
are required, which are achievable in practice. We believe that the
design philosophy presented here is generally valid in any traveling
wave antenna system in which the dispersion relation imposes a sharp
and tunable cut-off for each antenna arm. The figures of merit can
hence be expected to improve as new traveling wave antenna designs
are proposed in the field of plasmonics.

In closing, we have proposed a method to realize reconfigurable
plasmon antennas, e.g., for controlling the coupling of single
emitters with nodes in a quantum network.  Essential for our method
is the dispersion relation underlying traveling wave antennas that
provides a sharp tuneable cut-off. The specific design for a two
beam antenna presented in this paper uses host refractive index
changes from $n=1.5$ to $n=1.56$. Such changes are in the range
accessible with liquid crystals and phase change materials,
\cite{ElKallassi2007JOSAB,arnold2010APL} but above the level
accessible with, e.g., photochromic polymers \cite{Yager2006250} or
thermal index tuning. Particularly promising is the use of a photo
sensitive liquid crystal with potentially picosecond response time
to UV pulses. \cite{ElKallassi2007JOSAB,Yager2006250}  Embedding the
Yagi-Uda antennas inside a semiconductor matrix (Si or GaAs) would
allow fast and reversible switching using free carrier
excitation.\cite{EuserPRB2008} The operation wavelength in that case
shifts to the infrared due to the high host index. In addition to
the specific refractive index demands, we note several obstacles for
reconfigurable optical antennas. First we note that despite the high
directivity evident in Figs 2 and 3, the side lobes contain a
significant fraction of the far field emission.  Indeed, at an NA of
0.32 considered here, the two beams contain only about $50\%$ of the
emitted power (Fig. 3(e)).  Enlarging the NA, or embedding the
nanoscale antennas in micron scale dielectric waveguides will
suppress the side lobes, while retaining high light matter
interaction strength.  As a second obstacle, we note that turning
off a beam does not necessarily double  the brightness of the
remaining beam, as is evident from the drop in quantum efficiency in
Fig. 3(f). The  quantum efficiency is reduced because the branch
that is switched off still captures emission in the form of dark
plasmons. Such losses can be avoided by using other resonant
scatterers.  We have calculated that Yagi-Uda antennas also work
when made from high index (Si) particles. This configuration not
only avoids loss but would also allow easier switching, since the
particles themselves can be optically switched. Thirdly we notice
that Yagi-Uda antennas are limited by the fact that the dispersion
cut-off only occurs at one edge (blue edge). In multi-beam antennas
with more than 2 beams it is hence not possible to switch arbitrary
combinations of beams. Nonetheless, our design strategy  paves the
way for designing single emitter networks for quantum optics on the
chip.

We are especially indebted to Javier Garc\'{\i}a de Abajo for
providing MESME code. We  also gratefully acknowledge support from
Villum Fonden. This work is part of the research program of the
``Stichting voor Fundamenteel Onderzoek der Materie (FOM),'' which
is financially supported by the ``Nederlandse Organisatie voor
Wetenschappelijk Onderzoek (NWO).'' AFK thanks NWO-Vidi and
STW/Nanoned.

\end{document}